\documentclass[prl,showpacs,twocolumn,amsmath,amssymb]{revtex4-1}
\usepackage[utf8]{inputenc}

\usepackage{graphicx}
\usepackage{color}

\usepackage{graphicx}
\usepackage{color}
\usepackage{amsbsy}
\expandafter\let\csname equation*\endcsname\relax
\expandafter\let\csname endequation*\endcsname\relax 
\usepackage{amsmath}

\def\l{\lambda}

\def\beq{\begin{equation}}
\def\ee{\end{equation}}
\def\bi{\begin {itemize}}
\def\ei{\end{itemize}}

\def\lsim
{\protect \raisebox{-0.75ex}[-1.5ex]{$\;\stackrel{<}{\sim}\;$}}
\def\gsim
{\protect \raisebox{-0.75ex}[-1.5ex]{$\;\stackrel{>}{\sim}\;$}}
\def\lsimeq
{\protect \raisebox{-0.75ex}[-1.5ex]{$\;\stackrel{<}{\simeq}\;$}}
\def\gsimeq
{\protect \raisebox{-0.75ex}[-1.5ex]{$\;\stackrel{>}{\simeq}\;$}}

\def\l{\lambda}

\def\St{S_{\rm tot}(\tau)}

\def\t{_{\rm tot}}

\def\c{{\cal C}}

\def\beq{\begin{equation}}
\def\ee{\end{equation}}

\def\bi{\begin {itemize}}
\def\ei{\end{itemize}}

\def\F{{\cal F}}

\def\nn{\nonumber}

\begin{document}

\title{First and Second Law of Thermodynamics at strong coupling}
\author{
 Udo Seifert 
}

\affiliation{
{II.} Institut f\"ur Theoretische Physik, Universit\"at Stuttgart,
  70550 Stuttgart, Germany}
\begin{abstract}

For a  small driven system coupled strongly to a heat bath, 
internal energy and exchanged heat are identified such that they obey
the usual additive form of the first law. By identifying this exchanged heat
with the entropy change of the bath, the total entropy production
is shown to obey an integral fluctuation theorem on the trajectory level
implying  the
second law in the form of a Clausius inequalilty
on the ensemble level. In this Hamiltonian approach, 
the assumption of an initially uncorrelated state 
is not required. The conditions under which the proposed identification of 
heat is unique and experimentally accessible are clarified.

\end{abstract}

\maketitle

\def\lsim
{\protect \raisebox{-0.75ex}[-1.5ex]{$\;\stackrel{<}{\sim}\;$}}

\def\gsim
{\protect \raisebox{-0.75ex}[-1.5ex]{$\;\stackrel{>}{\sim}\;$}}

\def\lsimeq
{\protect \raisebox{-0.75ex}[-1.5ex]{$\;\stackrel{<}{\simeq}\;$}}

\def\gsimeq
{\protect \raisebox{-0.75ex}[-1.5ex]{$\;\stackrel{>}{\simeq}\;$}}

\def\Ek{E^{\rm {k}}}
\def\Ec{E^{\rm c}}
\def\Ech{\hat E^{\rm c}}
\def\Ekh{{{\hat E}^{\rm k}}}
\def\Ekhs{{{\hat E}^{\rm k'}}}
\def\Eh{\hat E}
\def\Ok{\Omega^{\rm k}}
\def\Oc{\Omega^{\rm c}}
\def\bk{\beta^{\rm k}}
\def\bc{\beta^{\rm c}}
\def\bks{\beta^{\rm k'}}
\def\bcs{\beta^{\rm c'}}

\def\k{^{\rm k}}
\def\c{^{\rm c}}

\def\dek{(\Delta \Ek)^2}

\def\ph{p^{\rm hist}}

\def\pb{{\bf p}}
\def\qb{{\bf q}}

\def\tt{_{\rm tot}}
\def\sy{_{\rm s}}

\def\xib{{\boldsymbol \xi}}
\def\H{{\cal H}}


\def\call{}
\def\W{{\cal  W}}
\def\E{{\cal E}}
\def\Q{{\cal  Q}}
\def\St{{\cal  S}_{\rm tot}}
\def\Ss{{\cal  S}}
\def\Sm{{\cal  S}_{\rm b}}
\def\st{s_{\rm tot}}

The thermodynamic analysis of a system coupled to a work source and
a heat bath of temperature $T$  typically relies crucially on the assumption that the coupling
to the bath is weak. If this condition is not met, partitioning the work
$\W$ spent in a process into dissipated heat $\Q$ and an increase in
internal energy of the system $\Delta\E$ in the form of a first law  
\beq
\W=\Delta \E + \Q
\label{eq:one}
\ee
leaves the question open whether at all, and, if so, in which form the interaction between 
system and bath is contained in the two terms on the right hand side.
The same issue arises in the second law when it is written in the 
additive form as a Clausius inequality
\beq
\Delta \St = \Delta \Ss +\Q/T\geq 0
\label{eq:two}
\ee splitting the total entropy change $\Delta  \St$ in one of the system $\Delta \Ss$ and one
of the bath given by the heat divided by $T$. 

Work is arguably the least problematic of the five quantities appearing above since it can easily be identified
 even in the presence of strong coupling.
By treating system and bath including the interaction as one big closed system that evolves
under a time-dependent Hamiltonian, the change of the latter from an initial state to a final one
represents work. If the initial state is drawn from a canonical ensemble for the whole
system, work is  known to obey exact relations like the
Jarzynski equality \cite{jarz97} and the Crooks relation \cite{croo99} 
 even in the presence of strong coupling 
\cite{jarz04} as, {\sl inter alia}, many 
single molecule experiments over the last decade have demonstrated convincingly
\cite{liph02,coll05,harr07a,rito08,gupt11,alem12}.
Since typical work values of even a few 
hundred $k_BT$ become tiny when divided by the number of molecules in the solution 
in contact with the bio-molecule, the change in
 the interaction between bath
and molecule is not necessarily negligible
compared to those in internal energy
of the molecule. The success of these experiments therefore rests
partially on the fact that
their interpretation does not require splitting the work into internal energy and heat for these
 strongly coupled
system.  On the other hand, for driven solid state devices, recent progess in ultra-sensitive calorimetry 
should soon make
  heat exchange directly accessible experimentally
\cite{viis15,camp15}. 

Exploring the role of strong coupling for {\sl equilibrium} thermodynamics has a
long history going back, in the classical case, at least to Kirkwood's concept
of a potential of mean force \cite{kirk35,roux99}, see, e.g.,  Ref. {\cite{geli09}
for a recent analysis. For quantum systems, the role of strong coupling has been
discussed in particular in the context of damped harmonic oscillators for quite some time
\cite{grab84,ford85} with a recent emphasis on apparent anomalies like a negative specific heat 
\cite{ingo09}. How to formulate a consistent thermodynamics 
for a 
strongly coupled system
under {\sl non-equilibrium} conditions, like relaxation after an initial quench or
genuine time-dependent driving, has found more attention lately for quantum systems than for
classical ones. Various approaches and schemes are discussed
\cite{nieu02,hoer08,camp09,kim10,espo10f,deff11,hilt11,camp11,kosl13,pucc13,gall14,anke14,carr15,chia15,hang15,espo15}
without arguably reaching a consensus yet on how to identify, beyond work,  the terms in
(\ref{eq:one}) and (\ref{eq:two}) uniquely.

Crucial aspects surface similarly in both frameworks, classical and quantum.
One common subtle issue concerns entropy production
since treating
the full system as closed, which works so nicely for an identification of work, implies
on the other hand that
the total change of  Gibbs, or Shannon, entropy  (classically),
or of the von Neumann entropy in the
quantum case, remains strictly constant
even under time-dependent driving. A positive entropy production results, however, if one 
ignores the correlations between system and bath, see, e.g.,  Ref. \cite{espo10f}.
Even then, however, the identification of heat is not  unique as,
e.g.,  the comparison of two
schemes for a simple relaxation for quantum Brownian motion has shown
\cite{pucc13}. Moreover, in these approaches,
one often assumes that initially  system and bath are individually equilibrated as if there
was no interaction. For most bio-molecular 
systems in aqueous solution, however, such an assumption is certainly rather unrealistic.

In this paper, we present an approach that allows to 
identify  the terms appearing
in the additive forms of the first and the second law consistently
for driven classical 
systems strongly coupled to a
heat bath without requiring an initially uncorrelated state.
In the limit of weak coupling, these quantities will become the established ones.
 A particular virtue of
this approach is that the terms appearing in (\ref{eq:one}) and (\ref{eq:two}) can be inferred
from measurements involving only observables of the system.

\def\X{{\cal X}}
\def \E{{\cal E}}
\def\F{{\cal F}}
\def\S{{\cal S}}
\def\bxi{\xi}
\def\Ht{H_{\rm tot}(\bxi, \l)}

\def\Hs{H_{\rm s}(\bxi_{\rm s}, \l)}
\def\Hm{H_{\rm b}(\bxi_{\rm b})}

\def\Hmt{H_{\rm b}(\bxi_{\rm b}^t)}

\def\Hi{H_{\rm i}(\bxi)}

\def\Hst{H_{\rm s}(\bxi_{\rm s}^\tau, \l^\tau)}
\def\Hit{H_{\rm i}(\bxi^\tau)}
\def\Hitt{H_{\rm i}(\bxi^t)}
\def\dHi{\delta H_{\rm i}(\bxi)}
\def\bHi{\bar H_{\rm i}(\bxi_{\rm s})}
\def\bHm{\bar H_{\rm b}(\bxi_{\rm b})}
\def\Hse{{\cal {H}}(\bxi_{\rm s},\l)}
\def\Use{{E}(\bxi_{\rm s},\l)}
\def\Uset{{E}(\bxi_{\rm s}^\tau,\l)}
\def\Hseta{{\cal {H}}(\tilde \bxi_{\rm s},\l)}
\def\Hset{{E}(\bxi_{\rm s}^t)}
\def\Hsen{{E}(\bxi_{\rm s}^0)}
\def\Ee{{\cal  E}}
\def\Eone{{\cal E}_1(\bxi_{\rm s},\l)}
\def\Fm{\F_{\rm b}}
\def\eq{^{\rm eq}}
\def\ssy{s}
\def\sm{s_{\rm b}}

\def\Bhi{B_{\rm i}(\bxi)}

As reference for the driven case, and to establish notation,
we first recall the equilibrium situation, see, e.g., \cite{geli09,camp11}. For
a system coupled to a heat bath,  the total Hamiltonian reads
\beq
\Ht=\Hs+\Hm+\Hi ,
\ee
comprising, in this order, system, bath,  and interaction Hamiltonian.
A micro state in the full phase space is written as $\bxi\equiv(\bxi_{\rm s},\bxi_{\rm b})$ where $\bxi_{\rm s}$ 
and $\bxi_{\rm b}$ denote micro states in the phase space of system and bath, respectively.
The control parameter $\l$, which will later be used to drive the system,
does neither affect the bath nor the interaction part of the Hamiltonian.
While in a Hamiltonian approach it may look more natural to consider a microcanonical equilibrium for
the full system, for technical reasons that will become clear later 
we rather choose a canonical equilibrium for the total
system at inverse temperature $\beta$. Then the probability to find 
the system part in a state $\bxi_{\rm s}$ is
given by 
\beq
p\eq(\bxi_{\rm s}|\l)=\exp[-\beta (\Hse - \F(\l))] .
\label{eq:peq}
\ee
Here, 
\beq
\Hse\equiv \Hs - \beta^{-1}\ln\langle \exp[-\beta \Hi] \rangle_{\rm b}.
\label{eq:Hse}
\ee
is an effective Hamiltonian, or,
in the jargon of physical
chemistry, a potential of mean force. It involves a canonical average over the pure 
bath (at fixed $\bxi_{\rm s}$) denoted
in the following 
by 
\beq
\langle ...\rangle_{\rm b}\equiv \int d\bxi_{\rm b} ... \exp[-\beta(\Hm-\Fm)],
\ee where $\Fm$ is the free energy of the pure bath.
 The $\l$-dependent
free energy of the system is defined through
\beq
\exp[-\beta \F(\l)]\equiv \int d\bxi_{\rm s} \exp[-\beta \Hse].
\ee 
Still in equilibrium, this free energy implies through the standard relation
$\Ss=\beta^2 \partial_\beta \F$  for the entropy  of the system 
\beq
\Ss(\l)=\int d\bxi_{\rm s}~p\eq(\bxi_{\rm s}) [-\ln p\eq(\bxi_{\rm s}) + \beta^2 \partial_\beta\Hse],
\label{eq:seq}
\ee setting Boltzmann's constant  to 1 throughout. Likewise, the internal energy
$\Ee=\F+\Ss/\beta$ becomes
\beq
\Ee(\l)=\int d\bxi_{\rm s}~p\eq(\bxi_{\rm s})\Use
\label{eq:eintt}
\ee
with
\beq
\Use\equiv 
\Hse +  \beta \partial_\beta\Hse .
\label{eq:eint}
\ee 
 In the weak coupling limit,  the three energy functions
$H_s,\H$, and $E$ converge.

The additional contribution $\sim \partial_\beta \Hse$
beyond what one might have expected
naively for entropy and internal energy takes into account that due to the finite interaction the bath is correlated
with the microstate $\bxi_{\rm s}$ of the system. In fact, with the standard canonical equilibrium for the total
system obeying in obvious notation the relation $\F\t= \E\t - \S\t/\beta$ and that for
the {\sl pure} bath with $\F_{\rm b}= \E_{\rm b}-\S_{\rm b}/\beta$ the above identified thermodynamic quantities of the
system fulfill
\beq
\X = \X\t - \X_{\rm b}
\ee
for $\X=\F,\E,\S$. This additive relation  indicates that in this approach the interaction is fully accounted for through modification of the
quantities refering to the system.

We now drive the system for a time $t$ through a time-dependent control parameter $\l^\tau$, with $0\leq \tau\leq t$.
The total system comprising the system proper, the heat bath and the interaction
 is assumed to be closed.
An initial phase point $\bxi^0$ then evolves in time deterministically
into $\bxi^t$. The corresponding mapping $\bxi^t=\bxi^t(\bxi^0)$ has Jacobian 1 due to
 Liouville's theorem. We first keep  a trajectory-based approach \cite {seif12,seki10}
in which all quantities become a function of the initial phase point $\bxi^0$.

 The work spent in the driving is  the total energy difference
\beq
w(\bxi^0)
\equiv \Delta \Ht \equiv H_{\rm tot}(\bxi^t,\l^t) -  H_{\rm tot}(\bxi^0,\l^0) .
\label{eq:work-zero}
\ee
Here, and in the following, $\Delta$ operating on a quantity implies
the difference of this quantity between final and initial value.
 Hamiltonian dynamics implies that this work can 
also be written as
\beq
w(\bxi^0) =\int_0^t d\tau \partial_\l\Hst \partial_\tau\l 
\label{eq:work}
\ee
which is the form used in stochastic energetics  \cite{seif12,seki10}. 
In fact, one could replace here $\partial_\l H_{\rm s}$ by either $\partial_\l \cal H$ 
or $\partial_\l E$ without changing 
the subsequent results since all three derivatives are the same.

As a key step in the present approach, motivated by (\ref{eq:eintt}),  internal energy of the system 
along a driven trajectory $\bxi^\tau$ is identified 
as $\Uset$, independent of the specific (and in any case
unknown) value of the instantaneous bath coordinates $\bxi_{\rm b}^\tau$. As we will show below, thus a consistent thermodynamic scheme arises.
This assignment of internal energy implies the
identification of dissipated 
heat as
\beq
q(\bxi^0)= w(\bxi^0)-\Delta \Use=\Delta [\Ht-\Use] .
\label{eq:heat}
\ee

It is instructive to show more explicitly how the interaction modifies the standard forms
of the terms in the first law. Writing 
\beq
\Hi= \langle \Hi \rangle_{\rm b} + \dHi
\ee
we separate the mean interaction, at fixed system coordinate $\bxi_{\rm s}$,
from its fluctuations $\dHi$. Similarly, the energy
of the bath is split according to
\beq
\Hm = \langle \Hm\rangle_{\rm b} + \delta\Hm .
\ee With (\ref{eq:Hse}) and (\ref{eq:eint}), 
the change in internal energy then becomes after little algebra
\begin{eqnarray}
\Delta \Use&=& \Delta[ \Hs + \langle \Hi \rangle_{\rm b} +\\
&~& +  \langle \dHi \Bhi\rangle_{\rm b} + \langle \delta\Hm\Bhi\rangle_{\rm b}] \nn
\end{eqnarray} where
\beq
\Bhi\equiv \exp[-\beta\dHi]/\langle \exp[-\beta\dHi]\rangle_{\rm b} .
\ee Thus the  average interaction is fully attributed to the internal energy, which,
however, also picks up two more contributions from the fluctuations. Correspondingly,
the heat (\ref{eq:heat}) becomes 
\begin{eqnarray}
q(\bxi^0)&=&\Delta [\Hm +\dHi \\&~& - \langle \dHi \Bhi\rangle_{\rm b} - \langle \delta\Hm\Bhi\rangle_{\rm b}] \nn
\label{eq:heat-two} .
\end{eqnarray}
 Beyond the standard expression of dissipated heat, which is the change in energy of the
bath $\Delta \Hm$, the first two  further
 contributions depend on how much the interaction fluctuates for a fixed system state
$\bxi_{\rm s}$. The last contribution depends on correlations of the interaction with fluctuations
of the bath.
In the weak coupling limit, these additional contributions vanish since the interaction becomes negligible.

 The first law is thus obeyed on the trajectory level by construction. It will remain valid
on the ensemble level after averaging with, in principle, any initial distribution $p^0(\bxi)$.
As physically sensible initial distributions we will choose from now on
\beq
p^0(\bxi)=p^0_1(\bxi_{\rm s})p_2\eq(\bxi_{\rm b}|\bxi_{\rm s},\l^0)
\label{eq:pini}
\ee
where
\beq
p_2\eq(\bxi_{\rm b}|\bxi_{\rm s},\l) \equiv \frac{\exp[-\beta(\Hi + \Hm-\Fm)]}{
\langle \exp[-\beta\Hi]\rangle_{\rm b}} .
\ee
The initial distribution of the system $p_1^0(\bxi_{\rm s})$ is arbitrary. For technical reasons,
we require that it does not vanish anywhere
on the phase space of the system.
The bath is assumed to be equilibrated initially for any  system state
$\bxi_{\rm s}$. In the following, averages with this initial distribution will
 be denoted
by $\langle ...\rangle$. Note that with
the option of an initially non-equilibrated system part
relaxation towards equilibrium at constant control parameter, e.g., after
a quench of the system, is covered by this framework as well. If $p^0_1(\bxi_{\rm s})$
is the equilibrium distribution (\ref{eq:peq}), then the initial
distribution (\ref{eq:pini}) corresponds to the canonical
equilibrium in the full phase space.

We now turn to checking the consistency of the proposed identification of heat with 
the additive form of the second law.
As a technical tool, we will use the trivial but powerful identity, or integral
fluctuation theorem (IFT),
\beq
1=\int d\bxi^t \rho(\bxi^t)=\langle \exp[\ln[\rho(\bxi^t(\xi^0))/p^0(\bxi^0)]]\rangle .
\label{eq:IFT}
\ee Liouville's theorem ensures that 
this IFT is valid for any normalized function $\rho(\bxi)$ provided the initial distribution
$p^0(\bxi)$  vanishes nowhere on the full phase space.
 By choosing
the legitimate factorized form 
\beq
\rho(\xi)=p_1^t(\bxi_{\rm s})p_2\eq(\bxi_{\rm b}|\bxi_{\rm s},\l^t)
\label{eq:rho}
\ee
where $p_1^\tau(\bxi_{\rm s})$ is the true marginal  distribution for $\bxi_{\rm s}$ at time $\tau$,
the IFT (\ref{eq:IFT}) becomes after trivial algebra
\beq
1=\langle \exp[-(\Delta \ssy(\bxi^0) + \beta q(\bxi^0))]\rangle 
\label{eq:IFTs}
\ee
where the average is over the initial distribution (\ref{eq:pini}). Here, the change in system entropy along the
trajectory is
\beq
\Delta \ssy (\bxi^0) \equiv -\ln p_1^t(\bxi_{\rm s}^t)+\ln p_1^0(\bxi_{\rm s}^0)
+ \Delta \beta^2\partial_\beta \Hse .
\label{eq:ssy} 
\ee
The first two terms amount to the change in stochastic entropy  
 familiar from stochastic thermodynamics \cite{seif05a}. The third contribution, called
intrinsic entropy in a related context \cite{seif11}, has the same physical origin as 
discussed above in equilibrium. 
If we now identify, as usual, the entropy change of the bath on the trajectory level 
 with the exchanged heat (times $\beta$), 
the exponent in (\ref{eq:IFTs}) becomes the total entropy production,
\beq
\Delta \st(\bxi^0)\equiv \Delta \ssy (\bxi^0) + \beta q(\bxi^0),
\ee
which thus  obeys  
 an IFT
\beq
\langle \exp[-\Delta \st(\bxi^0)]\rangle = 1 .
\label{eq:IFTt}
\ee
Even though this IFT looks like the one derived earlier using a stochastic
dynamics \cite{seif05a}, one should note that here it follows from a
Hamiltonian dynamics for a strongly coupled driven system.

The second law (\ref{eq:two}) for the calligraphic capitalized quantities
that denote the averages with respect to the initial distribution (\ref{eq:pini}) follows trivially
from Jensen's inequality applied to (\ref{eq:IFTt}). 
 On a mathematical level, we have thus shown that if  internal energy, heat, and the two
contributions to total entropy production are identified as suggested here, the additive
form of the first and second law are
valid in the presence of strong coupling.

Can heat and the other quantities be measured in an experiment
where one has access to the trajectory of the degrees of freedom of the system $\bxi_{\rm s}^\tau$ but, of course, 
 not to
the bath coordinates?
 Equilibration at fixed $\l$ yields $\Hse$ from  measuring the
corresponding
equilibrium distribution (\ref{eq:peq}). 
Repeating these measurements at a slightly different temperature
will lead to $\partial_\beta \Hse$ and thus to the internal energy
$\Use$ through ({\ref{eq:eint}). For the driven system, the work is accessible from observing the trajectory
$\bxi_{\rm s}^\tau$ using (\ref{eq:work}) since the 
$\l$-dependence of $\Hs$ is controlled in
an experiment. Hence, the heat can be inferred from 
evaluating (\ref{eq:heat}). Finally, the change in system entropy follows from measuring the marginal
distributions $p_1^t(\bxi_{\rm s})$ and $p_1^0(\bxi_{\rm s})$. Thus, all quantities are, at least in principle, measurable
experimentally from trajectories $\bxi_{\rm s}^\tau$ without ever having to measure a bath
degree of freedom. The ensemble quantities appearing in (\ref{eq:one}) and (\ref{eq:two}) then follow from averaging
the trajectory-resolved measurements.

A few further aspects, implications and perspectives are worth noting.
First, 
is this assignment of heat, or, equivalently, internal energy unique? On a formal level, there seems to be freedom.
Replacing internal energy, heat and change in system entropy on the trajectory level according to
$\Use\to\Use + \chi(\bxi_{\rm s},\l), q\to q- \Delta  \chi(\bxi_{\rm s},\l)$, and $\Delta \ssy\to \Delta \ssy +  
\beta\Delta \chi(\bxi_{\rm s},\l)$,
respectively, with an arbitrary system state function $ \chi(\bxi_{\rm s},\l)$, which vanishes in the weak coupling limit,
 leaves the first law (\ref{eq:heat}) and the IFT (\ref{eq:IFTt}) invariant. In fact, the choice $ \chi(\bxi_{\rm s},\l)=\Hse - \Use$ amounts to 
what has been discussed in Ref. \cite{pucc13} under the 
label "poised". The crucial point, however, is that any choice
$ \chi(\bxi_{\rm s},\l)\not=0$ will spoil the thermodynamic relation $\Ss=\beta^2\partial_\beta \F$, or, equivalently,
$d\E_{|\lambda}=Td\Ss_{|\lambda}$,  when applied on the ensemble level
to equilibrium. As long as one requires these latter relations for assigning the label "thermodynamically consistent" only the
present scheme with $ \chi(\bxi_{\rm s},\l)\equiv 0$ fulfills this criterion.

Second, we have assumed that the bath is in a system-state dependent equilibrium initially.
This choice is  physically
sensible  if there is a separation of time-scales between system and bath. Even with such a separation, 
however, the Hamiltonian dynamics will not precisely lead to a distribution of the form (\ref{eq:rho})
at time $t$.
Using the latter in (\ref{eq:IFT}) should therefore be interpreted as a mathematical convenience
for deriving the IFT (\ref{eq:IFTs}) rather than as a statement about the true distribution. As an aside, note that substituting 
the canonical
distribution of the full system at $\l^t$ for $\rho(\xi)$ into (\ref{eq:IFT}) yields the strongly coupled Jarzynski equality
\cite{jarz04} for an initially equilibrated system in one line. 

Third, equality in the second law usually requires a quasistatic process. In our approach, the second
law (\ref{eq:two}) follows from the IFT (\ref{eq:IFTt}). Any IFT requires for a saturation of the corresponding
inequality that the underlying
distribution for the exponent is delta-like. Thus equality in (\ref{eq:two}) holds if and only if 
$\Delta \st(\bxi^0)$ vanishes identically for all initial micro states $\bxi^0$. Ultimately, this requirement implies
that the distribution for the full system starts and remains canonical throughout the process.
In this respect, the strong coupling case does not differ from weak coupling. In fact, from a more 
physical perspective, one would expect that a  moderate or strong coupling should 
facilitate equilibration and hence the realization of quasistatic conditions
even more than the common idealized weak coupling case does.

Fourth, so far, we have not split the total volume into one of the system and one of the bath
which would give rise to a pressure term. It would be interesting to explore which modifications 
arise from such a perspective in the case of strong coupling
\footnote{ C. Jarzynski, talk presented at the FQMT2015 in Prague.}.

Finally, since the main part of this paper dealt with classical systems, it is worth emphasizing that the
present scheme suggests, by analogy with (\ref{eq:eint}), as an internal energy operator for the system in the quantum case 
\beq
\hat E \equiv -(1+\beta\partial_\beta)[\beta^{-1}\ln {\rm Tr}_{\rm b}\exp[-\beta(\hat H_{\rm s}+\hat H_{\rm i}+ \hat H_{\rm b}-F_{\rm b})]],
\ee where hats denote operators, the trace is over the bath degrees of freedom 
and $\exp(-\beta F_{\rm b})\equiv {\rm Tr}_{\rm b} \exp(-\beta \hat H_{\rm b})$. In general, this operator $\hat E$ will be
a quite complicated function of temperature and the parameters of the total Hamiltonian. The change in internal
energy then follows, in principle, from two point measurements of $\hat E$ at $\tau=0$ and $\tau = t$. 
Since, in general,  $[\hat E,\hat H_{\rm tot}]\not = 0$, 
 work as given by the difference in total energy can not be measured simultaneously. Hence, heat as the
difference of work and internal energy is not accessible through this route  in the quantum case.

In conclusion, for a classical driven system strongly coupled to a heat bath not only work but also
internal energy, dissipated heat
and entropy production can be identified on the level of a trajectory of the 
system. Total entropy production obeys an integral fluctuation theorem implying, on the ensemble level, a 
consistent interpretation of the second law as a Clausius inequality. For an experimental
realization, the heat accompanying conformational changes of mechanically manipulated bio-molecules 
should be accessible experimentally through measurements at two different temperatures as suggested here. 
While the theory is not confined to this particular class, with such experiments these molecules could turn out to
become one paradigm for studying heat exchange in small driven strongly coupled systems.

Acknowledgments: I thank S. Goldt and P. Pietzonka for a critical reading of this manuscript.

\bibliographystyle{aipnum4-1}
%

\end{document}